\documentclass[aps,prl,superscriptaddress,preprint,tightenlines,floatfix]{revtex4}

\usepackage[dvips]{graphicx}
\usepackage[dvips]{color}
\usepackage{epsfig}
\usepackage{amssymb,amsmath}
\usepackage{lscape}

\newcommand{\bdk}{$B^{\pm}\to DK^{\pm}$}
\newcommand{\bdtk}{$B^{\pm}\to \tilde{D}K^{\pm}$}

\newcommand{\bdsk}{$B^{\pm}\to D^{*}K^{\pm}$}
\newcommand{\bdstk}{$B^{\pm}\to \tilde{D}^{*}K^{\pm}$}

\newcommand{\bddsk}{$B^{\pm}\to D^{(*)}K^{\pm}$}
\newcommand{\bddstk}{$B^{\pm}\to \tilde{D}^{(*)}K^{\pm}$}

\newcommand{\bdpi}{$B^{\pm}\to D\pi^{\pm}$}
\newcommand{\bdtpi}{$B^{\pm}\to \tilde{D}\pi^{\pm}$}

\newcommand{\bdspi}{$B^{\pm}\to D^{*}\pi^{\pm}$}
\newcommand{\bdstpi}{$B^{\pm}\to \tilde{D}^{*}\pi^{\pm}$}

\newcommand{\bddspi}{$B^{\pm}\to D^{(*)}\pi^{\pm}$}
\newcommand{\bddstpi}{$B^{\pm}\to \tilde{D}^{(*)}\pi^{\pm}$}

\newcommand{\dsdpi}{$D^{*\pm}\to D\pi^{\pm}$}
\newcommand{\dsdpis}{$D^{*\pm}\to D\pi_s^{\pm}$}

\newcommand{\dkpp}{$\bar{D^0}\to K_S\pi^+\pi^-$}
\newcommand{\dtkpp}{$\tilde{D}\to K_S\pi^+\pi^-$}

\begin{document}

\preprint{BELLE-CONF-0476}

\title{Measurement of \boldmath{$\phi_3$} with Dalitz Plot
Analysis of \boldmath{\bddsk} Decay at Belle}
\affiliation{Aomori University, Aomori}
\affiliation{Budker Institute of Nuclear Physics, Novosibirsk}
\affiliation{Chiba University, Chiba}
\affiliation{Chonnam National University, Kwangju}
\affiliation{Chuo University, Tokyo}
\affiliation{University of Cincinnati, Cincinnati, Ohio 45221}
\affiliation{University of Frankfurt, Frankfurt}
\affiliation{Gyeongsang National University, Chinju}
\affiliation{University of Hawaii, Honolulu, Hawaii 96822}
\affiliation{High Energy Accelerator Research Organization (KEK), Tsukuba}
\affiliation{Hiroshima Institute of Technology, Hiroshima}
\affiliation{Institute of High Energy Physics, Chinese Academy of Sciences, Beijing}
\affiliation{Institute of High Energy Physics, Vienna}
\affiliation{Institute for Theoretical and Experimental Physics, Moscow}
\affiliation{J. Stefan Institute, Ljubljana}
\affiliation{Kanagawa University, Yokohama}
\affiliation{Korea University, Seoul}
\affiliation{Kyoto University, Kyoto}
\affiliation{Kyungpook National University, Taegu}
\affiliation{Swiss Federal Institute of Technology of Lausanne, EPFL, Lausanne}
\affiliation{University of Ljubljana, Ljubljana}
\affiliation{University of Maribor, Maribor}
\affiliation{University of Melbourne, Victoria}
\affiliation{Nagoya University, Nagoya}
\affiliation{Nara Women's University, Nara}
\affiliation{National Central University, Chung-li}
\affiliation{National Kaohsiung Normal University, Kaohsiung}
\affiliation{National United University, Miao Li}
\affiliation{Department of Physics, National Taiwan University, Taipei}
\affiliation{H. Niewodniczanski Institute of Nuclear Physics, Krakow}
\affiliation{Nihon Dental College, Niigata}
\affiliation{Niigata University, Niigata}
\affiliation{Osaka City University, Osaka}
\affiliation{Osaka University, Osaka}
\affiliation{Panjab University, Chandigarh}
\affiliation{Peking University, Beijing}
\affiliation{Princeton University, Princeton, New Jersey 08545}
\affiliation{RIKEN BNL Research Center, Upton, New York 11973}
\affiliation{Saga University, Saga}
\affiliation{University of Science and Technology of China, Hefei}
\affiliation{Seoul National University, Seoul}
\affiliation{Sungkyunkwan University, Suwon}
\affiliation{University of Sydney, Sydney NSW}
\affiliation{Tata Institute of Fundamental Research, Bombay}
\affiliation{Toho University, Funabashi}
\affiliation{Tohoku Gakuin University, Tagajo}
\affiliation{Tohoku University, Sendai}
\affiliation{Department of Physics, University of Tokyo, Tokyo}
\affiliation{Tokyo Institute of Technology, Tokyo}
\affiliation{Tokyo Metropolitan University, Tokyo}
\affiliation{Tokyo University of Agriculture and Technology, Tokyo}
\affiliation{Toyama National College of Maritime Technology, Toyama}
\affiliation{University of Tsukuba, Tsukuba}
\affiliation{Utkal University, Bhubaneswer}
\affiliation{Virginia Polytechnic Institute and State University, Blacksburg, Virginia 24061}
\affiliation{Yonsei University, Seoul}
  \author{K.~Abe}\affiliation{High Energy Accelerator Research Organization (KEK), Tsukuba} 
  \author{K.~Abe}\affiliation{Tohoku Gakuin University, Tagajo} 
  \author{N.~Abe}\affiliation{Tokyo Institute of Technology, Tokyo} 
  \author{I.~Adachi}\affiliation{High Energy Accelerator Research Organization (KEK), Tsukuba} 
  \author{H.~Aihara}\affiliation{Department of Physics, University of Tokyo, Tokyo} 
  \author{M.~Akatsu}\affiliation{Nagoya University, Nagoya} 
  \author{Y.~Asano}\affiliation{University of Tsukuba, Tsukuba} 
  \author{T.~Aso}\affiliation{Toyama National College of Maritime Technology, Toyama} 
  \author{V.~Aulchenko}\affiliation{Budker Institute of Nuclear Physics, Novosibirsk} 
  \author{T.~Aushev}\affiliation{Institute for Theoretical and Experimental Physics, Moscow} 
% ...
\collaboration{The Belle Collaboration}

\date{\today}

\begin{abstract} 

We present a measurement of the unitarity triangle angle $\phi_3$
using a Dalitz plot analysis of the three-body decay of the neutral 
$D$ meson from the \bddsk\ process. 
Using a 253 fb$^{-1}$ data sample collected by the Belle experiment,  
we obtain 276 signal candidates for \bdk\ and 
69 candidates for \bdsk, where the neutral $D$ meson decays into 
$K_S \pi^+ \pi^-$.
From a combined maximum likelihood fit to the \bdk\ and \bdsk\ modes, 
we obtain $\phi_3=68^{\circ}\;^{+14^{\circ}}_{-15^{\circ}}
\mbox{(stat)}\pm 13^{\circ}
\mbox{(syst)}\pm 11^{\circ}(\mbox{model})$. The
corresponding two standard deviation 
interval is $22^{\circ}<\phi_3<113^{\circ}$. 
\end{abstract}
\maketitle

\section{Introduction}

\begin{figure}[!bht]
  \begin{center}
  \epsfig{figure=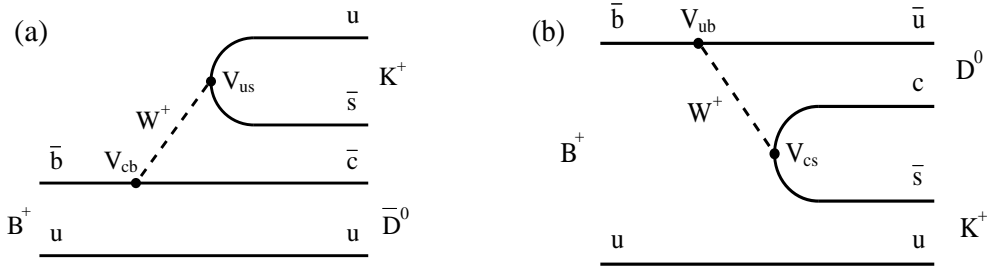,width=0.8\textwidth}
  \caption{Feynman diagrams of (a) dominant $B^+\to \bar{D^0}K^+$ and 
           (b) suppressed $B^+\to D^0K^+$ decays}
  \label{diags}
  \end{center}
\end{figure}

Determinations of the Cabbibo-Kobayashi-Maskawa
(CKM) \cite{ckm} matrix elements provide important checks on
the consistency of the Standard Model and ways to search
for new physics.  Various methods using $CP$ violation in $B\to D K$ decays have been
proposed \cite{glw,dunietz,eilam,ads} to measure the unitarity triangle
angle $\phi_3$. These methods are based on two key observations:
neutral $D^{0}$ and $\bar{D^0}$
mesons can decay to a common final state, and the decay
$B^+\to D^{(*)} K^+$ can produce neutral $D$ mesons of both flavors
via $\bar{b}\to \bar{c}u\bar{s}$ (Fig.~\ref{diags}a)
and $\bar{b}\to \bar{u}c\bar{s}$ (Fig.~\ref{diags}b) transitions,
with a relative phase $\theta_+$ between the two interfering
amplitudes that is the sum, $\delta + \phi_3$, of strong and weak interaction
phases.  For the charge conjugate mode, the relative phase is
$\theta_-=\delta-\phi_3$. 

Recently, three body final states common to $D^0$ and
$\bar{D^0}$, such as $K_S\pi^+\pi^-$ \cite{giri}, were suggested as 
promising modes for the extraction of $\phi_3$. 
In the Wolfenstein parameterization of the CKM matrix elements, 
the amplitudes of the two diagrams
that contribute to the decay $B^+\to D K^+$
are given by $M_1\sim V_{cb}^*V_{us}\sim A\lambda^3$ 
(for the $\bar{D^0} K^+$ final state) and
$M_2\sim V_{ub}^*V_{cs}\sim A\lambda^3(\rho+i\eta)$
(for $D^0 K^+$).
The two amplitudes 
$M_1$ and $M_2$ interfere as the $D^0$ and $\bar{D^0}$ mesons decay
into the same final state $K_S \pi^+ \pi^-$; 
we denote the admixed state as $\tilde{D}$. Assuming no $CP$ 
asymmetry in neutral $D$ decays, the amplitude of the $B^+$ decay 
can be written as
\begin{equation}
\label{intdist}
M_+=f(m^2_+, m^2_-)+re^{i\phi_3+i\delta}f(m^2_-, m^2_+) 
\end{equation}
and the corresponding amplitude for the charge conjugate $B^-$ decay is 
can be written as 
\begin{equation}
\label{intdist_m}
M_-=f(m^2_-, m^2_+)+re^{-i\phi_3+i\delta}f(m^2_+, m^2_-), 
\end{equation}
where $m_{+}^2$ and $m_{-}^2$ are the squared 
invariant masses of the $K_S \pi^+$ and
$K_S \pi^-$ combinations, respectively, and $f(m_+, m_-)$ 
is the complex amplitude for the \dkpp\ decay. 
The absolute value of the ratio between 
the two interfering amplitudes, $r$, is predicted to be 0.1--0.2. 

Once the functional form of $f$ is fixed by a \dkpp\ decay model, 
the $\tilde{D}$ Dalitz distributions for the $B^+$ and $B^-$ decays can be 
fitted simultaneously using the above expressions with $r$, $\phi_3$, 
and $\delta$ as free parameters.
The \dkpp\ decay model can be determined
from a large sample of flavor-tagged \dkpp\ decays 
produced in continuum $e^+e^-$ annihilation.

The current measurement is an update of our analysis \cite{belle_phi3}, 
which was based on 140 fb$^{-1}$ data sample. 
Both measurements are based on two decay modes, \bdtk\ and 
\bdstk\ ($D^*\to D\pi^0$). 
Recently, the same technique has been applied by the BABAR collaboration
\cite{babar_phi3} with consistent results. 

\section{Event selection}

We use a 253 fb$^{-1}$ data sample, corresponding to
$275\times 10^6$ $B\bar{B}$ pairs, collected by 
the Belle detector. The decays \bdk\ and 
\bdsk, $D^{*}\to D\pi^0$ are selected for the 
determination of $\phi_3$; the decays \bdpi\ and 
\bdspi\ with $D^{*}\to D\pi^0$ serve as control samples. 
We require the neutral $D$ meson to decay to the 
$K_S\pi^+\pi^-$ final state in all cases.
We also select decays of \dsdpi\ produced via the 
$e^+e^-\to c\bar{c}$ continuum process as a high-statistics 
sample to determine the \dkpp\ decay amplitude. 
The Belle detector is described in detail elsewhere \cite{belle}. 
The event selection procedures are the same as was described in the 
previous analysis~\cite{belle_phi3}.

We use $\Delta M =  M_{K_S\pi^+\pi^- \pi^+_s} - 
M_{K_S\pi^+\pi^-}$ and $M_{K_S\pi^+\pi^-}$ distributions to select 
the $D^{*\pm}\to D\pi^{\pm}_s$ events, where $\pi^{\pm}_s$ stands 
for ``slow pion'' that is a distinct signature for  this decay.  
The fit to the $\Delta M$ distribution yields $186854\pm 856$ signal 
events and $6126\pm 65$ background events in the signal region
($144.6\mbox{ MeV}/c^2<\Delta M<146.4\mbox{ MeV}/c^2$). 
The corresponding background fraction is 3.2\%.

\begin{figure}[!htb]
  \epsfig{figure=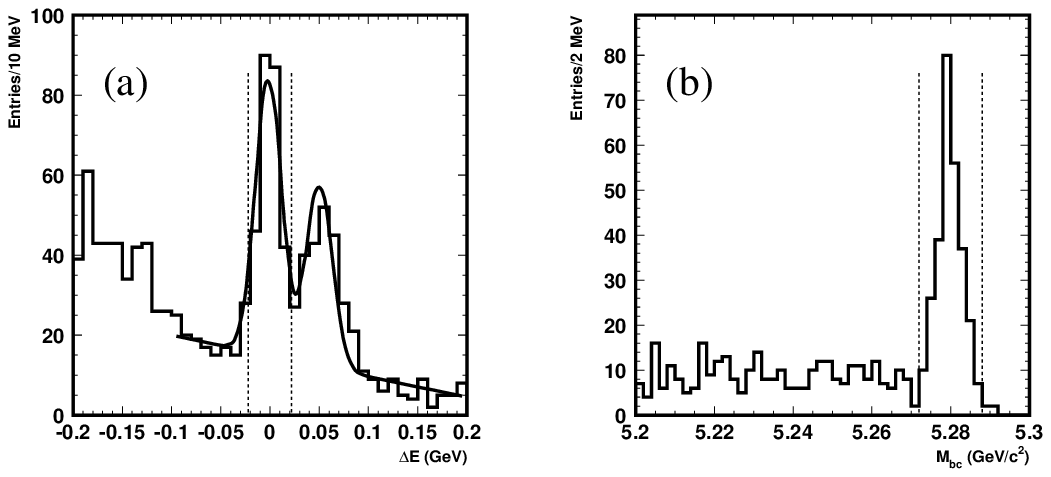,width=\textwidth}
  \caption{(a) $\Delta E$ and (b) $M_{\rm bc}$ distributions for the \bdk
  candidates. Dashed lines show the signal region. 
  The histogram shows the data; the smooth curve in (a) is the fit result.}
  \label{b2dk_sel}

  \epsfig{figure=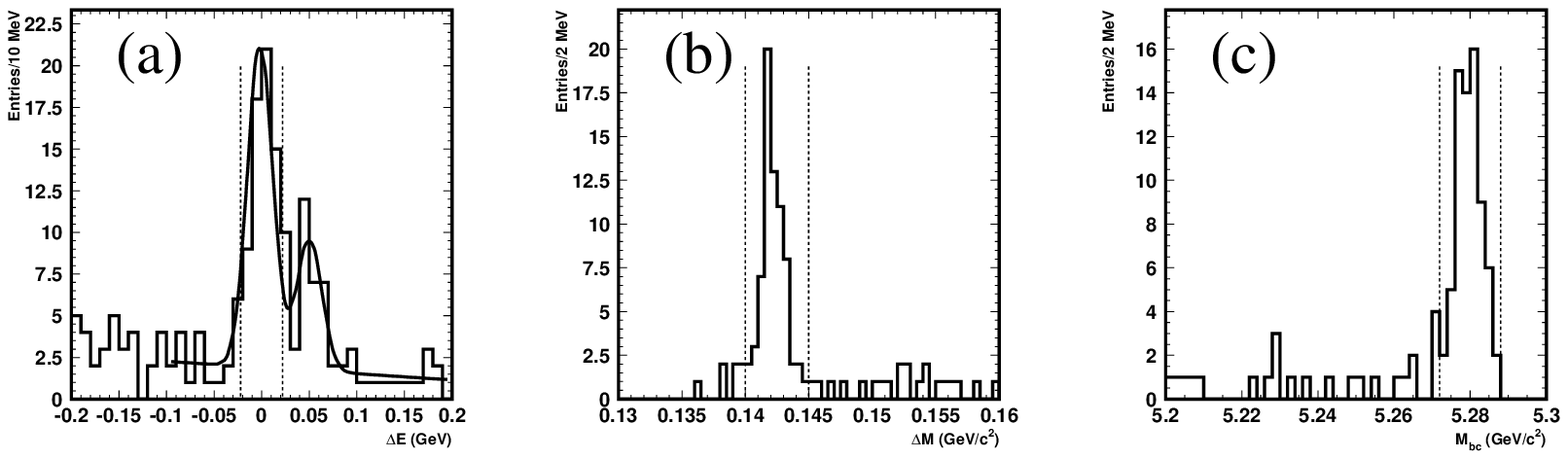,width=\textwidth}
  \caption{(a) $\Delta E$, (b) $\Delta M$ and (c) $M_{\rm bc}$ 
           distributions for the \bdsk candidates. 
           Dashed lines show the signal region. The histogram shows the data;
           the smooth curve in (a) is the fit result.}
  \label{b2dsk_sel}
\end{figure}

The selection of $B$ candidates is based on the CM energy difference
$\Delta E = \sum E_i - E_{\rm beam}$ and the beam-constrained $B$ meson mass
$M_{\rm bc} = \sqrt{E_{\rm beam}^2 - (\sum p_i)^2}$, where $E_{\rm beam}$ 
is the CM beam 
energy, and $E_i$ and $p_i$ are the CM energies and momenta of the
$B$ candidate decay products. The requirements for signal 
candidates are $5.272$~GeV/$c^2<M_{\rm bc}<5.288$ GeV/$c^2$ and $|\Delta E|<0.022$ GeV. 
The $\Delta E$ and $M_{\rm bc}$ distributions for \bdk\ candidates are
shown in Fig.~\ref{b2dk_sel}. The peak in the $\Delta E$ distribution at 
$\Delta E=50$ MeV is due to \bdpi\ decays, where the pion is misidentified
as a kaon.
The \bdk\ selection efficiency (11\%) is determined from 
a Monte Carlo (MC) simulation. The number of events passing all selection 
criteria is 276. 
The background fraction is determined from a binned fit to the $\Delta E$
distribution, in which the signal is represented by a Gaussian distribution 
with mean
$\Delta E=0$, the \bdpi\ component is represented by a Gaussian 
distribution with 
mean $\Delta E=50$ MeV and the remaining background is modeled by a 
linear function. The contributions in the signal region are found 
to be $209\pm 16$ signal events, $2.6\pm 0.3$ \bdpi\ events and 
$65\pm 5$ events in the linear background. 
The overall background fraction is $25\pm 2$\%.

Figure \ref{b2dsk_sel} shows the $\Delta E$, $M_{\rm bc}$ and $\Delta M$ 
distributions for \bdsk\ candidates. The selection 
efficiency is 6.2\%. The number of events satisfying 
the selection criteria is 69. The background fraction is determined in 
the same way as for \bdk\ events. The fit of the $\Delta E$
distribution yields $58\pm 8$ signal events, $8.3\pm 1.5$ events 
corresponding to the linear background and $0.44\pm 0.10$ \bdspi\ 
events in the signal region. The background fraction is $13\pm 2$\%.

\section{Determination of $\bar{D^0} \to K_S \pi^+ \pi^-$ decay model}

The amplitude $f$ of the \dkpp\ decay is represented
by a coherent sum of two-body decay amplitudes plus one non-resonant 
decay amplitude,
\begin{equation}
  f(m^2_+, m^2_-) = \sum\limits_{j=1}^{N} a_j e^{i\alpha_j}
  \mathcal{A}_j(m^2_+, m^2_-)+
    b e^{i\beta}, 
  \label{d0_model}
\end{equation}
where $N$ is the total number of resonances, 
$\mathcal{A}_j(m^2_+, m^2_-)$, $a_j$ and 
$\alpha_j$ are the matrix element, amplitude and phase, respectively, 
of the $j$-th resonance, and $b$ and $\beta$ are the amplitude
and phase of the non-resonant component. The total phase and amplitude 
are arbitrary. To be consistent with other analyses 
\cite{babar_phi3,dkpp_cleo}
we have chosen the $\bar{D^0}\to K_S\rho$ 
mode to have unit amplitude and zero relative phase. 
The description of the matrix elements follows Ref.~\cite{cleo_model}. 

For the $\bar{D^0}$ model we use a set of 18 two-body amplitudes. 
These include five Cabibbo-allowed amplitudes: $K^*(892)^+\pi^-$, 
$K^*(1410)^+\pi^-$, 
$K_0^*(1430)^+\pi^-$, 
$K_2^*(1430)^+\pi^-$ and $K^*(1680)^+\pi^-$;  
their doubly Cabibbo-suppressed partners; and eight amplitudes with
$K_S$ and a $\pi\pi$ resonance:
$K_S\rho$, $K_S\omega$, $K_Sf_0(980)$, $K_Sf_2(1270)$, 
$K_Sf_0(1370)$, $K_S\rho(1450)$, $K_S\sigma_1$ and $K_S\sigma_2$. 
The differences from our previous analysis \cite{belle_phi3} are, 
i) addition of  $K_S\rho(1450)$, $K^*(1410)^+\pi^-$ and its doubly 
   Cabibbo-suppressed mode, 
ii) use of the Gounaris-Sakurai \cite{gounaris} amplitude description for 
the $K_S\rho$ and $K_S\rho(1450)$ contributions, and 
iii) the mass and width for the $f_0(1370)$ state taken from \cite{aitala} 
($M=1434$ MeV$/c^2$, $\Gamma=173$ MeV$/c^2$).


We use an unbinned maximum likelihood technique to fit the Dalitz plot 
distribution to the model described by Eq.~\ref{d0_model}. 
We minimize the inverse 
logarithm of the likelihood function in the form
\begin{equation}
  -2 \log L = -2\left[\sum\limits^n_{i=1}\log p(m^2_{+,i}, m^2_{-,i}) - 
  \log\int\limits_D p(m^2_+, m^2_-)dm^2_+ d m^2_-\right], 
  \label{log_l}
\end{equation}
where $i$ runs over all selected event candidates, and
$m^2_{+,i}$, $m^2_{-,i}$ are measured Dalitz plot
variables. The integral in the second term accounts for the overall 
normalization of the probability density. 

The Dalitz plot density is represented by
\begin{equation}
  p(m^2_+, m^2_-) = \varepsilon(m^2_+, m^2_-)
  \int\limits^{\infty}_{-\infty}|M(m^2_++\mu^2, m^2_-+\mu^2)|^2
  \exp\left(-\frac{\mu^2}{2\sigma^2_{m}(m^2_{\pi\pi})}\right)d\mu^2
  +B(m^2_+, m^2_-),
  \label{density}
\end{equation}
where $M(m^2_+, m^2_-)=f(m^2_+, m^2_-)$ is the decay amplitude described 
by Eq.~\ref{d0_model}, 
$\varepsilon(m^2_+, m^2_-)$ is the efficiency, 
$B(m^2_+, m^2_-)$ is the background density, 
$\sigma_m(m^2_{\pi\pi})$ is the resolution of the squared invariant 
mass $m^2_{\pi\pi}$ of two pions
($m^2_{\pi\pi}=M^2_D+M^2_K+2M^2_{\pi}-m^2_+-m^2_-$). 
The free parameters of the minimization are the amplitudes
$a_j$ and phases $\alpha_j$ of the resonances (except for the $K_S\rho$
component, for which the parameters are fixed), 
the amplitude $b$ and phase $\beta$ of the non-resonant component
and the masses and widths of the $\sigma_1$ and $\sigma_2$ scalars.

\begin{figure}[!htb]	
  \begin{center}
  \epsfig{figure=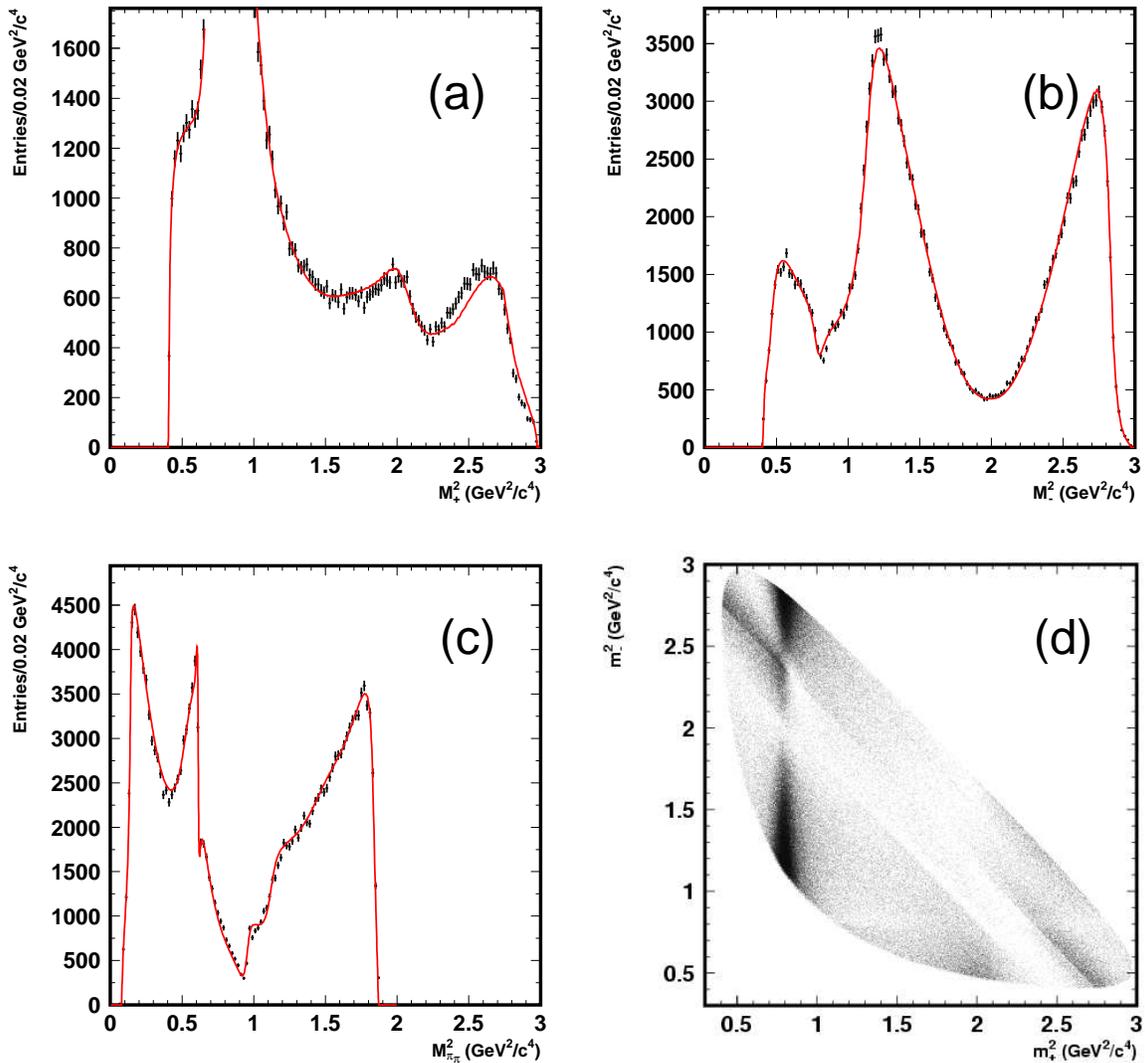,width=\textwidth}
  \caption{(a) $m^2_+$, (b) $m^2_-$, (c) $m^2_{\pi\pi}$
   distributions
   and (d) Dalitz plot for the \dkpp\ decay from the \dsdpis\ process.
   The points with error bars show the data, the smooth curve is the 
   fit result.}
  \label{ds2dpi_plot}
  \end{center}
\end{figure}

The procedures for determining the background
density, the efficiency, and the resolution of the squared invariant
mass, are the same as in the previous analysis. 
The \dkpp\ Dalitz plot distribution, as well as
its projections with the fit results superimposed, are shown in 
Fig.~\ref{ds2dpi_plot}. 
The fit results are given in Table~\ref{dkpp_table}. 
The parameters of the $\sigma$ resonances obtained 
in the fit are: 
$M_{\sigma_1}=520\pm 15$ MeV/$c^2$,
$\Gamma_{\sigma_1}=466\pm 31$ MeV/$c^2$, 
$M_{\sigma_2}=1059\pm 6$ MeV/$c^2$, and
$\Gamma_{\sigma_2}=59\pm 10$ MeV/$c^2$.
The large peak in the $m^2_+$ distribution 
corresponds to the dominant $\bar{D^0}\to K^*(892)^+\pi^-$ mode. 
The minimum in the $m^2_-$ distribution at 0.8~GeV$^2/c^4$
is due to destructive interference with the doubly Cabibbo 
suppressed $\bar{D^0}\to K^*(892)^-\pi^+$ amplitude. In the $m^2_{\pi\pi}$
distribution, the $\bar{D^0}\to K_S\rho$ contribution 
is visible around 0.5~GeV$^2/c^4$
with a steep edge on the upper side due to interference with 
$\bar{D^0}\to K_S\omega$. The minimum around 0.9~GeV$^2/c^4$ is due to 
the decay $\bar{D^0}\to K_S f_0(980)$ interfering destructively with
other modes.

\begin{table}[!htb]
\caption{Fit results for \dkpp\ decay. Errors are statistical only.
The results for the $\sigma_1$, $\sigma_2$ masses and widths are given 
in the text.}
\label{dkpp_table}
\begin{tabular}{|l|c|c|c|} \hline
Intermediate state           & Amplitude 
			     & Phase ($^{\circ}$) 
			     & Fit fraction
			     \\ \hline

$K_S \sigma_1$               & $1.57\pm 0.10$
			     & $214\pm 4$
			     & 9.8\%
                             \\

$K_S\rho^0$                  & $1.0$ (fixed)                                 
                             & 0 (fixed)
			     & 21.6\%
                             \\

$K_S\omega$                  & $0.0310\pm 0.0010$
			     & $113.4\pm 1.9$
			     & 0.4\%
                             \\

$K_S f_0(980)$               & $0.394\pm 0.006$
			     & $207\pm 3$
			     & 4.9\%
                             \\

$K_S \sigma_2$               & $0.23\pm 0.03$
			     & $210\pm 13$
			     & 0.6\%
                             \\

$K_S f_2(1270)$              & $1.32\pm 0.04$
			     & $348\pm 2$
			     & 1.5\%
                             \\

$K_S f_0(1370)$              & $1.25\pm 0.10$
			     & $69\pm 8$
			     & 1.1\%
                             \\

$K_S \rho^0(1450)$           & $0.89\pm 0.07$
			     & $1\pm 6$
			     & 0.4\%
                             \\

$K^*(892)^+\pi^-$            & $1.621\pm 0.010$
			     & $131.7\pm 0.5$
			     & 61.2\%
                             \\ 

$K^*(892)^-\pi^+$            & $0.154\pm 0.005$
			     & $317.7\pm 1.6$
			     & 0.55\%
                             \\

$K^*(1410)^+\pi^-$	     & $0.22\pm 0.04$
			     & $120\pm 14$
			     & 0.05\%
			     \\

$K^*(1410)^-\pi^+$	     & $0.35\pm 0.04$
			     & $253\pm 6$
			     & 0.14\%
			     \\

$K_0^*(1430)^+\pi^-$         & $2.15\pm 0.04$
			     & $348.7\pm 1.1$
			     & 7.4\%
                             \\

$K_0^*(1430)^-\pi^+$         & $0.52\pm 0.04$
			     & $89\pm 4$
			     & 0.43\%
                             \\

$K_2^*(1430)^+\pi^-$         & $1.11\pm 0.03$
			     & $320.5\pm 1.8$
			     & 2.2\%
                             \\

$K_2^*(1430)^-\pi^+$         & $0.23\pm 0.02$
			     & $263\pm 7$
			     & 0.09\%
                             \\

$K^*(1680)^+\pi^-$           & $2.34\pm 0.26$
			     & $110\pm 5$
			     & 0.36\%
                             \\

$K^*(1680)^-\pi^+$           & $1.3\pm 0.2$
			     & $87\pm 11$
			     & 0.11\%
                             \\

non-resonant                 & $3.8\pm 0.3$
			     & $157\pm 4$
			     & 9.7\%
                             \\ 
\hline
\end{tabular}
\end{table}

The unbinned likelihood technique does not provide a reliable 
criterion for the goodness of fit. To check the quality of the 
fit, we make use of the large number of events in our sample and
perform a binned $\chi^2$ test by dividing the Dalitz plot into 
square regions $0.05\times 0.05$ GeV$^2/c^4$. The test yields 
$\chi^2=2543$ for 1106 degrees of freedom. More detailed studies 
are required in order to understand the precise dynamics of 
\dkpp\ decay. However, for the purpose of measuring $\phi_3$, 
we take the fit discrepancy into account in the model uncertainty. 

\section{Dalitz plot analysis of $B^{\pm} \to D K^{\pm}$ decay}

\begin{figure}[!htb]
  \epsfig{figure=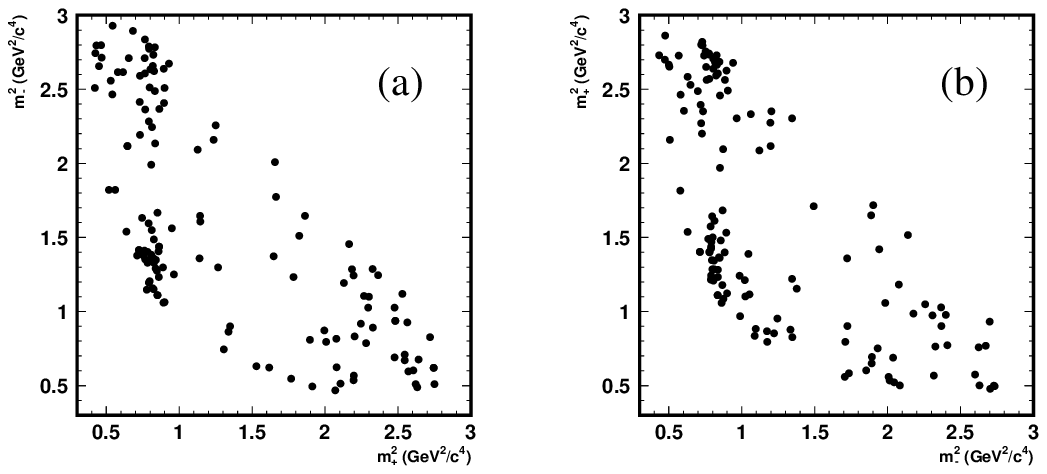,width=\textwidth}
  \caption{Dalitz plots of \dtkpp\ decay from (a) $B^+\to \tilde{D} K^+$
  and (b) $B^-\to \tilde{D} K^-$.}
  \label{b2dk_plots}

  \epsfig{figure=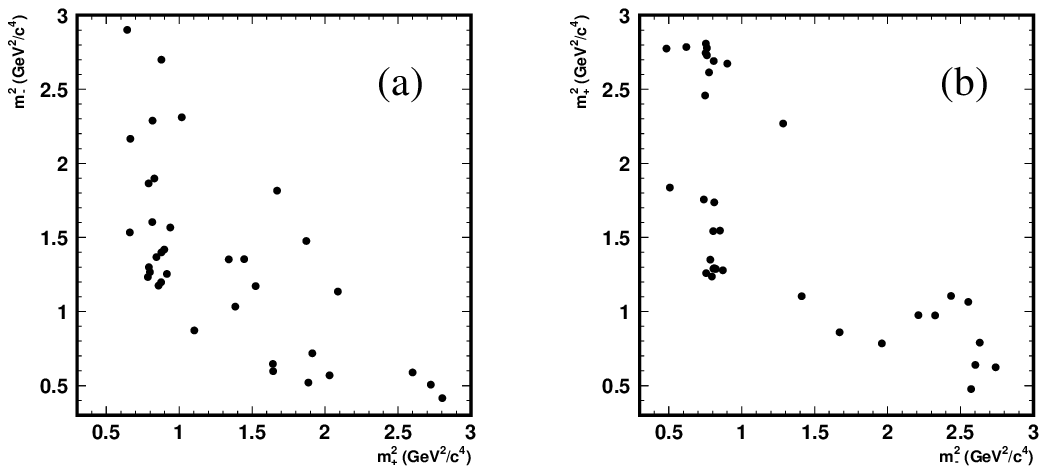,width=\textwidth}
  \caption{Dalitz plots of \dtkpp\ decay from (a) $B^+\to \tilde{D^{*}} K^+$
  and (b) $B^-\to \tilde{D^{*}} K^-$.}
  \label{b2dsk_plots}
\end{figure}

The Dalitz plot distributions for the \dtkpp\ decay are shown in 
Figs.~\ref{b2dk_plots} and \ref{b2dsk_plots}
for the \bdtk\ and \bdstk\ decays, respectively. 
These distributions are 
fitted by minimizing the combined logarithmic 
likelihood function 
\begin{equation}
  -2\log L=-2\log L_- -2\log L_+,
\end{equation}
where $L_-(L_+)$ are the likelihoods of $B^-(B^+)$ data given 
by Eq.~\ref{log_l}. The corresponding Dalitz plot densities 
$p_{\pm}(m_+^2, m_-^2)$ are given by Eq.~\ref{density} with 
decay amplitudes $M_{\pm}$ described by Eq.~\ref{intdist} ($B^+$ data)
and Eq.~\ref{intdist_m} ($B^-$ data). 
The $\bar{D^0}$ decay model $f$ is fixed, and the free parameters of the
fit are the amplitude ratio $r$ and phases $\phi_3$ and $\delta$. 



We consider five sources of background (see Table~\ref{bck_table}), and  
determine the fraction and Dalitz plot shape for each component. 
The largest contribution comes from two kinds of continuum events:
random combination of tracks, and correctly
reconstructed neutral $D$ mesons combined with random kaons.
We estimate their fractions to be $21.0\pm 1.7$\% for \bdk\ and 
$9.0\pm 2.2$\% for \bdsk\ using an event sample in which we 
make requirements that
primarily select continuum events but reject $B\bar{B}$ events. 
The shape of their Dalitz plot distribution is 
parameterized by a third-order polynomial in the variables $m^2_+$ and 
$m^2_-$ for the combinatorial background component and a sum of $D^0$ 
and $\bar{D^0}$ shapes for real neutral $D$ mesons combined with random kaons. 
 
\begin{table}
\caption{Fractions of different background sources.}
\label{bck_table}
\begin{tabular}{|l|c|c|} \hline
Background source                     & \bdk & \bdsk \\ \hline
$q\bar{q}$ combinatorial              & $21.0\pm 1.7$\% & $9.0\pm 2.2$\%    \\
$B\bar{B}$ events other than $B^{\pm}\to D^{(*)}K^{\pm}/\pi^{\pm}$ 
                                      & $2.3\pm 0.2$\%  & $3.1\pm 0.4$\%    \\
\bddspi with $K/\pi$ misID            & $0.9\pm 0.1$\%  & $0.7\pm 0.2$\%    \\
Combinatorics in $D^0$ decay          & $0.6\pm 0.1$\%  & $0.6\pm 0.1$\%    \\
Combinatorial kaon in $B^{\pm}\to D^{(*)}K^{\pm}$ decay & $<$0.4\% (95\% CL) & $<$0.4\% (95\% CL) \\ \hline
Total                                 & $25\pm 2$\%     & $13\pm 2$\%     \\ \hline
\end{tabular}
\end{table}

The background from $B\bar{B}$ events is subdivided into four
categories. 
The $D^{(*)} K^{\pm}$ and $D^{(*)} \pi^{\pm}$ combinations coming 
from the decay of $D^{(*)}$ 
from one $B$ meson and $K^{\pm}$ and $\pi^{\pm}$ from the other $B$ decay 
constitute the largest part of the $B\bar{B}$ background. 
We obtain their fractions of $2.3\pm 0.2$\% for \bdk\ and $3.1\pm 0.4$\%
for \bdsk\ using a MC study. Their Dalitz plot shapes are parameterized
by a second-order polynomial in the variables $m^2_+$ and $m^2_-$ for \bdk\ 
and by a linear function in $m^2_+$ and $m^2_-$ plus $D^0$ shape 
for \bdsk.
The background fraction from the process \bddspi\ with a pion misidentified 
as a kaon is obtained by fitting the $\Delta E$ 
distribution. The corresponding Dalitz plot shape is that of 
$\bar{D^0}$ without the opposite flavor admixture. 
The fractions for this background are $0.9\pm 0.1$\% 
for \bdk\ and $0.7\pm 0.2$\% for \bdsk.
The \bddsk\ events where one of the neutral $D$ meson decay products is 
combined with
a random kaon or pion were studied using a MC data set where 
one of the charged $B$ mesons from the $\Upsilon(4S)$ decays into the 
$D^{(*)}K$ state. 
The estimated background fraction is $0.6\pm 0.1$\% 
for both \bdk\ and \bdsk\ modes. The Dalitz plot shape is parameterized
by a linear function in the variables $m^2_+$ and $m^2_-$ 
plus a $D^0$ amplitude. 
Events in which a correctly reconstructed neutral $D$ is combined 
with a random charged kaon 
are of importance. A  half of the kaons will have 
the wrong sign and will be misinterpreted 
as decays of $D$ mesons of the opposite flavor, thus introducing distortion 
in the
most sensitive area of the Dalitz plot. In the MC sample, we find no events 
of this kind, which allows us to set an upper limit of 0.4\%
(at 95\% CL) on the fraction for this contribution.


To test the consistency of the fitting procedure, the same fitting 
procedure is 
applied to the \bddstpi\ control samples as to the \bddstk\ signal. 
In the case of \bddstpi, a small amplitude ratio is expected 
($r\sim |V_{ub} V^*_{cd}|/|V_{cb}V^*_{ud}|\sim 0.01-0.02$). 
Here, we consider $B^+$ and $B^-$ data separately, 
to check for the absence of $CP$ violation. 
The free parameters of the Dalitz plot fit are $r_{\pm}$ and 
$\theta_{\pm}$, where $\theta_{\pm}=\delta\pm\phi_3$. 
The fit results for \bdtpi\ sample (3425 events) are $r_+=0.039\pm 0.021$, 
$\theta_+=240^{\circ}\pm 28^{\circ}$
for $B^+$ data and 
$r_-=0.047\pm 0.018$, 
$\theta_-=193^{\circ}\pm 24^{\circ}$ for $B^-$ data. 
It should be noted that, since the value of $r$ is positive definite, 
the error of this parameter does not serve as a good 
measure of the $r \simeq 0$ hypothesis. To demonstrate the deviation of the 
amplitude ratio $r$ from zero, the real and imaginary parts of the complex 
amplitude ratio $r e^{i\theta}$ are more suitable. 
Figure~\ref{test_constr} (a) shows the complex 
amplitude ratio constraints for the $B^+$ and $B^-$ data separately. 
It can be seen that the amplitude ratios differ from zero by roughly two
standard deviations for both cases but are not inconsistent with  
the expected value of $r\sim 0.01-0.02$.  

The other control sample, \bdstpi\ with $\tilde{D}^{*}$ decaying to 
$\tilde{D}\pi^0$, also does not show any significant deviation from 
$r\simeq 0$. 
The results of the fit to the \bdstpi\ sample (642 events) are 
$r_+=0.015\pm 0.042$, $\theta_+=169^{\circ}\pm 186^{\circ}$, 
$r_-=0.086\pm 0.049$, $\theta_-=280^{\circ}\pm 30^{\circ}$ and 
are shown in Fig.~\ref{test_constr} (b).

\section{Results}

\begin{figure}[!htb]
  \epsfig{figure=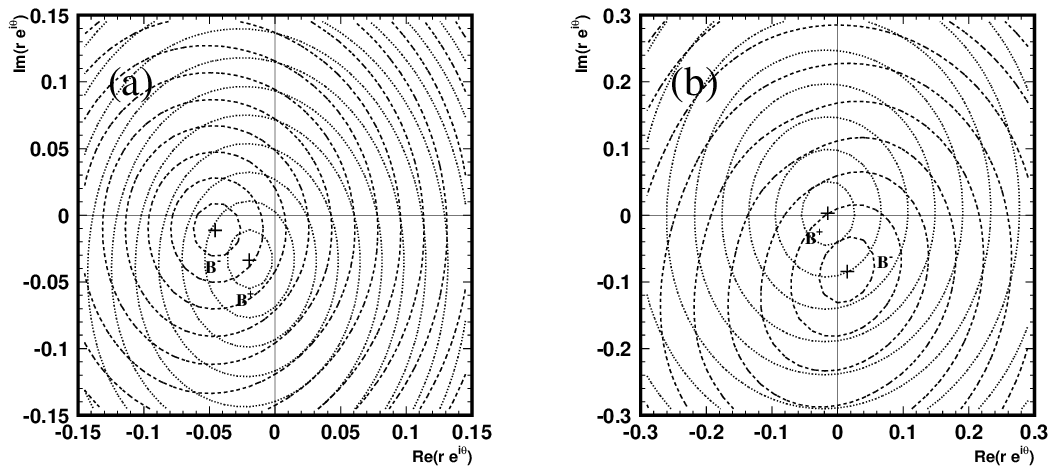,width=\textwidth}
  \caption{Constraint plots of the complex amplitude ratio
           $re^{i\theta}$ for (a) \bdtpi\ and
           (b) \bdstpi\ decays.
	   Contours indicate integer multiples of the standard deviation.
           Dotted contours are from $B^+$ data, dashed 
           contours are from $B^-$ data.}
  \label{test_constr}
  \epsfig{figure=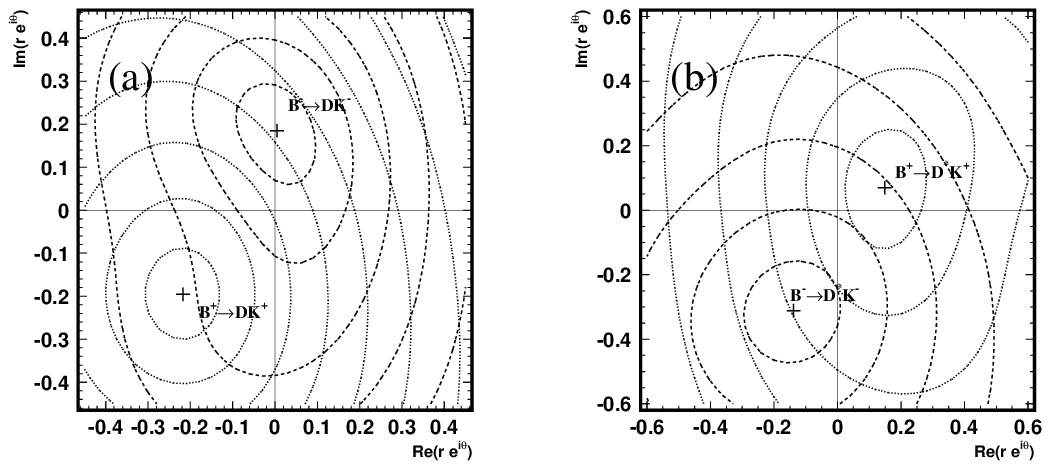,width=\textwidth}
  \caption{Constraint plots of the complex amplitude ratio
           $re^{i\theta}$ for (a) \bdtk\ and (b) \bdstk\ decays. 
	   Contours indicate integer multiples of the standard deviation.
           Dotted contours are from $B^+$ data, dashed 
           contours are from $B^-$ data.}
  \label{compl_constr}
\end{figure}

Fig.~\ref{compl_constr} shows the constraints on the complex 
amplitude ratio $r e^{i\theta}$ for the \bdtk\ and \bdstk\ samples. 
It can be seen that in both signal samples a significant
non-zero value of $r$ is observed.
A difference between the phases $\theta_+$ and $\theta_-$ 
is also apparent in both the \bdtk\ and \bdstk\ samples, 
which indicates a deviation of $\phi_3$ from zero. 

A combined unbinned maximum likelihood fit to the 
$B^+$ and $B^-$ samples with 
$r$, $\phi_3$ and $\delta$ as free parameters yields the following values: 
$r=0.25\pm 0.07$, $\phi_3=64^{\circ}\pm 15^{\circ}$, 
$\delta=157^{\circ}\pm 16^{\circ}$ for the \bdtk\ sample and 
$r=0.25\pm 0.12$, $\phi_3=75^{\circ}\pm 25^{\circ}$, 
$\delta=321^{\circ}\pm 25^{\circ}$ for the \bdstk\ sample.
The errors quoted here are obtained from the likelihood fit.
These errors are a good representation of the statistical uncertainties for
a Gaussian likelihood distribution, however in our case
the distributions are highly non-Gaussian. In addition, the errors
for the strong and weak phases depend on the values of the
amplitude ratio $r$ ({\it e.g.} for $r=0$ there is 
no sensitivity to the phases). A more reliable estimate of the
statistical uncertainties is obtained using a large number
of MC pseudo-experiments as discussed below.


The model used for the \dkpp\ decay is one of the main sources of 
systematic error for our analysis. Since the Dalitz density 
is proportional to the absolute value squared of the decay amplitude, 
the phase $\phi(m^2_+, m^2_-)$ of the complex amplitude is not 
directly measured. The phase variations across 
the Dalitz plot distribution are therefore the result of model assumptions. 

A MC simulation is used for estimating the effects of the model uncertainties. 
Event samples are generated according to the Dalitz plot distribution 
described by the amplitude given by Eq.~\ref{intdist} 
with the resonance parameters extracted from 
our fit of continuum $D^0$ data. We then fit the distribution using  
different models for $f(m_+, m_-)$ (see Table~\ref{model_table}). 
We scan the phases $\phi_3$
and $\delta$ in their physical regions and take the maximum 
deviations of the fit parameters ($(\Delta r)_{\rm max}$, 
$(\Delta\phi_3)_{\rm max}$, and $(\Delta\delta)_{\rm max}$) 
as model uncertainty estimates. 
The values for $(\Delta r)_{\rm max}$, 
$(\Delta\phi_3)_{\rm max}$ and $(\Delta\delta)_{\rm max}$
quoted in Table~\ref{model_table} are obtained with 
the value $r=0.13$. For larger $r$ values, the model uncertainty 
tends to be smaller, 
so our estimate of the model uncertainty is conservative. 

All the fit models are based on Breit-Wigner parameterizations 
of resonances. Since a Breit-Wigner amplitude can only describe narrow resonances well, 
the usual technique to deal with broad states is to introduce 
Blatt-Weisskopf form factors for the $\bar{D^0}$ meson ($F_D$) and intermediate 
resonance ($F_r$) and a $q^2$-dependence of the resonance width $\Gamma$. 
These quantities have substantial theoretical uncertainties. 
We therefore use a fit without Blatt-Weisskopf form factors 
and with a constant resonance width to estimate such an error.
We also use a model containing only narrow resonances
($K^*(892)$, $\rho$, doubly Cabibbo-suppressed $K^*(892)$ and $f^0(980)$)
with the wide ones approximated by the flat non-resonant term.
The study of the model errors is summarized in Table~\ref{model_table}.
Our estimate of the systematic uncertainty on $\phi_3$ is $11^{\circ}$.

\begin{table}
\begin{center}
\caption{Estimation of model uncertainty.}
\vspace{0.5\baselineskip}
\label{model_table}
\begin{tabular}{|c|c|c|c|c|} \hline
Fit model & $(\Delta r)_{\rm max}$ 
          & $(\Delta\phi_3)_{\rm max}$ ($^{\circ}$) 
          & $(\Delta\delta)_{\rm max}$ ($^{\circ}$) \\
\hline
$F_r=F_D=1$
    & 0.01 & 3.1 & 3.3 \\
$\Gamma(q^2)=Const$
    & 0.02 & 4.7 & 9.0 \\
Narrow resonances plus non-resonant term
    & 0.03 & 9.9 & 18.2 \\ \hline
Total
    & 0.04 & 11  & 21  \\
\hline
\end{tabular}
\end{center}
\end{table}


There are other potential sources of systematic
error such as uncertainties in the background Dalitz density, 
efficiency variations over the phase space, $m^2_{\pi\pi}$ resolution, 
and possible fit biases.
These are listed in Table~\ref{syst_table} for the \bdtk\ 
and \bdstk\ modes separately. The effect of background Dalitz density 
is estimated by extracting the background shape parameters  
from the $M_D$ sidebands and by using a flat background distribution.
The maximum deviation of the fit parameters from the ``standard''
background parameterization is assigned as the corresponding
systematic error. The effect of the uncertainty in the background 
fraction is studied by varying the background
fraction by one standard deviation.
The efficiency shape and $m^2_{\pi\pi}$ resolution 
are extracted from the MC simulations. 
To estimate their contributions to the systematic error, we 
repeat the fit using a flat efficiency and a fit model that
does not take the resolution into account, respectively. 
The biases due to the efficiency shape differ for \bdtk\ and \bdstk\ 
samples, but since we expect the values of the efficiency systematics
to be close for the two modes, we assign the maximum value
of the bias as the corresponding systematic error. 

\begin{table}
\caption{Contributions to the experimental systematic error.}
\label{syst_table}
\begin{tabular}{|l|c|c|c|c|c|c|} \hline
                     & \multicolumn{3}{|c|}{\bdtk}
                     & \multicolumn{3}{|c|}{\bdstk} \\ 
		     \cline{2-7}
Source               & $\Delta r$ & $\Delta\phi_3$ ($^{\circ}$) & $\Delta\delta$ ($^{\circ}$) 
                     & $\Delta r$ & $\Delta\phi_3$ ($^{\circ}$) & $\Delta\delta$ ($^{\circ}$) \\ \hline
Background shape     & 0.027      & 5.7                  & 4.1       
                     & 0.014      & 3.1                  & 5.3                  \\
Background fraction  & 0.006      & 0.2                  & 1.0     
                     & 0.005      & 0.7                  & 1.4                  \\
Efficiency shape     & 0.012      & 4.9                  & 2.4    
                     & 0.002      & 3.5                  & 1.0                  \\
$m^2_{\pi\pi}$ resolution  & 0.002      & 0.3                  & 0.3   
                     & 0.002      & 1.7                  & 1.4                  \\
Control sample bias  & 0.004      & 10.2                 & 10.2   
                     & 0.004      & 9.9                  & 9.9                  \\
Total                & 0.03      & 13                 & 11
                     & 0.02      & 11                 & 11                 \\ \hline
\end{tabular}
\end{table}


We use a frequentist technique to evaluate the 
statistical significance of the measurements. 
To obtain the  probability density function (PDF) of the fitted 
parameters as a function of the true parameters, which is needed for this 
method, we employ a ``toy" MC technique that uses a
simplified MC simulation of the experiment which incorporates
the same efficiencies, resolution and backgrounds as
used in the data fit.  This MC is used
to generate several hundred experiments for a given set of
$r$, $\theta_+$ and $\theta_-$ values. For each simulated
experiment, Dalitz plot distributions are generated
with equal numbers of events as in the data, 137 and 139 events 
for $B^-$ and $B^+$ decays, correspondingly, for \bdtk\ mode and 
34 and 35 events for $B^-$ and $B^+$ for
\bdstk\ mode. The simulated Dalitz plot distributions
are subjected to the same fitting procedure that is applied
to the data. This is repeated for different values of $r$,
producing distributions of the fitted parameters that
are used to produce a functional form of the PDFs of the 
reconstructed values for any set of input parameters.

We parameterize the PDF of a set of 
fit parameters $(r,\phi_3,\delta)$, assuming 
the errors of parameters 
${\rm Re}(r_{\pm} e^{i\theta_{\pm}})$ and 
${\rm Im}(r_{\pm} e^{i\theta_{\pm}})$ are 
uncorrelated and have Gaussian distributions with equal RMS
which we denote as $\sigma$. 
The PDF of the parameters $(r_{\pm}, \theta_{\pm})$ for the true 
parameters $(\bar{r}_{\pm}, \bar{\theta}_{\pm})$ is thus written as
\begin{equation}
  d^2 P(r_{\pm},\theta_{\pm}| \bar{r}_{\pm},\bar{\theta}_{\pm})=
       \frac{1}{2\pi\sigma^2}\exp\left[-\frac{
       (r_{\pm}\cos\theta_{\pm}-\bar{r}\cos\bar{\theta}_{\pm})^2+
       (r_{\pm}\sin\theta_{\pm}-\bar{r}\sin\bar{\theta}_{\pm})^2
       }{2\sigma^2}\right] r_{\pm}dr_{\pm}d\theta_{\pm}. 
\end{equation}
To obtain the PDF for the parameters $(r,\phi_3,\delta)$ we fix
$r=r_+=r_-$ and substitute the total phases with $\delta+\phi_3$ and 
$\delta-\phi_3$:
\begin{equation}
  \frac{d^3 P}{dr d\phi_3 d\delta}(r,\phi_3,\delta| 
                       \bar{r}, \bar{\phi}_3, \bar{\delta})=
    \frac{d^2 P}{dr_+ d\theta_+}(r,\delta+\phi_3| \bar{r},
                                 \bar{\delta}+\bar{\phi}_3) 
    \frac{d^2 P}{dr_- d\theta_-}(r,\delta-\phi_3| \bar{r}, 
                                 \bar{\delta}-\bar{\phi}_3). 
  \label{fit_pdf}
\end{equation}
There is only one free parameter $\sigma$ which is obtained
from the unbinned maximum likelihood fit of the MC distribution 
to Eq.~\ref{fit_pdf}. 
The value of $\sigma$ is equal to 0.10 for the \bdtk\ decay and 
$\sigma=0.18$ for the \bdstk\ decay. 

\begin{figure}
  \epsfig{figure=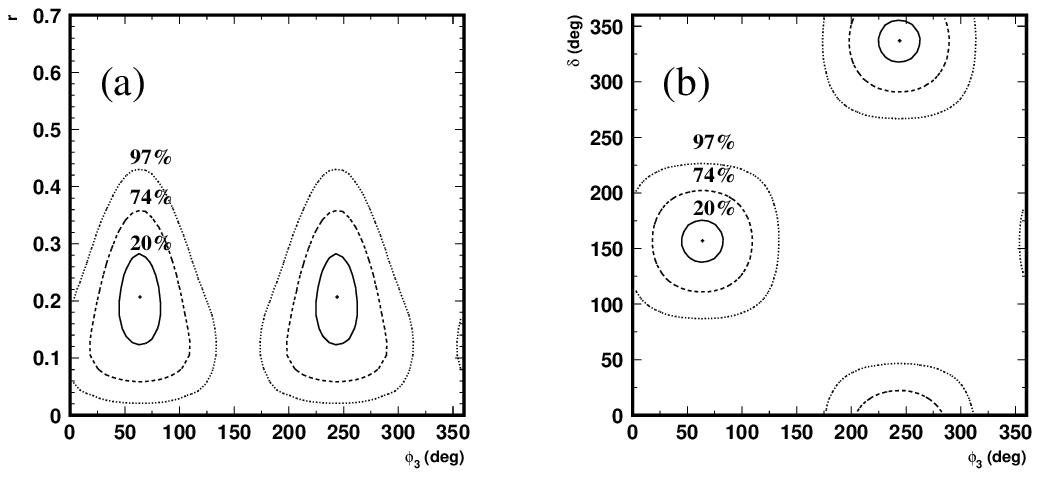,width=\textwidth}
  \caption{Confidence regions for the pairs of parameters (a) ($r$, $\phi_3$) 
           and (b) ($\phi_3, \delta$) for the \bdtk\ sample.}
  \label{b2dk_neum}

  \epsfig{figure=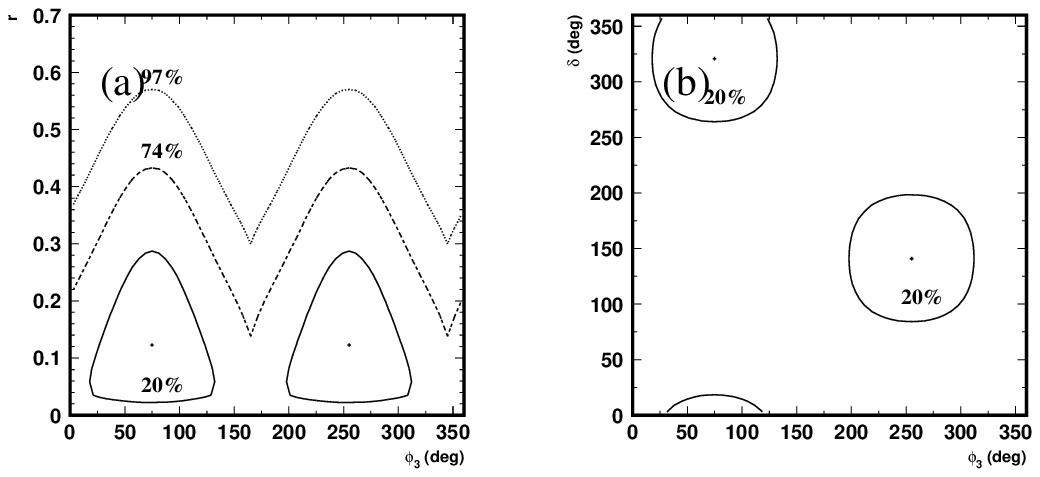,width=\textwidth}
  \caption{Confidence regions for the pairs of parameters (a) ($r$, $\phi_3$)
           and (b) ($\phi_3, \delta$) for the \bdstk\ sample.}
  \label{b2dsk_neum}
\end{figure}

Once the PDF is obtained, we can calculate the confidence 
level $\alpha$ for each set of true parameters 
$(\bar{r},\bar{\phi}_3,\bar{\delta})$ 
for given measurements, 
$(r, \phi_3, \delta) = (0.25, 64^{\circ}, 157^{\circ})$ for 
the $\tilde{D} K^\pm$ mode and 
$(0.25, 75^{\circ}, 321^{\circ})$ for the $\tilde{D^*} K^\pm$ mode. 
The confidence regions for the pairs of parameters 
$(\phi_3, \delta)$ and $(\phi_3, r)$ are shown in Fig.~\ref{b2dk_neum} 
(\bdtk\ mode) and Fig.~\ref{b2dsk_neum} (\bdstk\ mode).
They are the projections of the corresponding 
confidence regions in the three-dimensional parameter space. 
We show the 20\%, 74\% and 97\% confidence level regions, 
which correspond to 
one, two, and three standard deviations for a three-dimensional Gaussian
distribution.

For the final results, we use the central values that  are obtained by 
maximizing the PDF and the statistical errors corresponding to the 20\% 
confidence region (one standard deviation). Of the two possible 
solutions ($\phi_3$, $\delta$ and $\phi_3+180^{\circ}$, $\delta+180^{\circ}$) 
we choose the one with $0<\phi_3<180^{\circ}$. The final results are 
\begin{equation} 
r = 0.21 \pm 0.08 \pm 0.03 \pm 0.04,
~\phi_3=64^{\circ} \pm 19^{\circ} \pm 13^{\circ} \pm 11^{\circ},~ 
\delta=157^{\circ} \pm 19^{\circ} \pm 11^{\circ} \pm 21^{\circ}
\end{equation}
for the \bdtk\ mode and 
\begin{equation}
r = 0.12^{+0.16}_{-0.11} \pm 0.02 \pm 0.04,
~\phi_3=75^{\circ} \pm 57^{\circ} \pm 11^{\circ} \pm 11^{\circ},
~\delta=321^{\circ} \pm 57^{\circ} \pm 11^{\circ} \pm 21^{\circ}
\end{equation}
for the \bdstk\ mode.  The first, second, and third errors are
statistical, systematic, and model dependent errors.

While the $\phi_3$ and $\delta$ values that are determined
from the toy MC are consistent with those that are determined in the unbinned 
maximum likelihood fits for both $\tilde{D} K^\pm$ and $\tilde{D^*}
K^\pm$, the corresponding $r$ values are significantly different. This
is due to a bias in the unbinned maximum likelihood. Since $r$ is a
positive-definite quantity, the fit tends to return a larger value for
$r$ than its true value, particularly when $r$ is small. 

In the frequentist approach, the significance of $CP$ violation 
is evaluated by finding the confidence level for the most probable 
$CP$ conserving point, {\it i.e.} the point with $r=0$ or $\phi_3=0$, 
for which the confidence level $\alpha(\bar{r},\bar{\phi}_3,\bar{\delta})$
is minimal. This procedure gives $\alpha=94$\% for the \bdtk\ sample
and $\alpha=38$\% for \bdstk.

The two events samples, \bdk\ and \bdsk, are combined 
in order to obtain a more accurate measurement of $\phi_3$. 
The technique we use to obtain the combined measurement is 
also based on a frequentist approach. 
Here we have five true parameters ($\bar{\phi_3}$, $\bar{r}_1$, $\bar{r}_2$, 
$\bar{\delta}_1$ and $\bar{\delta}_2$, where the indices 1 and 2
correspond to \bdk\ and \bdsk\ modes, respectively) 
and six reconstructed parameters ($r$, $\phi_3$ and $\delta$ 
for each of the two modes). 
The PDF for the reconstructed parameters is written as
\begin{equation}
  \frac{dP}{dx}(x,\mu)=
  \frac{d^3 P_{B\to D^0K}   }{dr d\phi_3 d\delta}
  (r_1,(\phi_3)_1,\delta_1|\bar{r_1},\bar{\phi}_3,\bar{\delta_1})
  \frac{d^3 P_{B\to D^{*0}K}}{dr d\phi_3 d\delta}
  (r_2,(\phi_3)_2,\delta_2|\bar{r_2},\bar{\phi}_3,\bar{\delta_2}), 
\end{equation}
where $x=(dr_1, d(\phi_3)_1, d\delta_1, dr_2, d(\phi_3)_2, d\delta_2)$ is a 
vector of the reconstructed parameters, and
$\mu=(\bar{\phi_3}, \bar{r}_1, \bar{r}_2, 
\bar{\delta}_1, \bar{\delta}_2)$ is a vector of the true parameters. 
Using this PDF and Feldman-Cousins likelihood ratio ordering \cite{feldman},
 we can calculate the confidence level $\alpha (\mu)$. 
%
This approach gives $\phi_3 = 68^{\circ}$ as the central value.
The one standard deviation interval for $\phi_3$ 
(which corresponds to the 3.7\% confidence level for the case of 
a five-dimensional Gaussian distribution) is 
$\phi_3=68^{\circ}\;^{+14^{\circ}}_{-15^{\circ}}$. 
 
Since the \bdk\ contribution dominates in the combined measurement, we 
use its value of the systematic uncertainty, which is $13^{\circ}$, as
an estimate of the systematic uncertainty in the combined $\phi_3$ 
measurement. 
The $\phi_3$ result from the combined analysis is 
\begin{equation}
\phi_3=68^{\circ}\;^{+14^{\circ}}_{-15^{\circ}}\pm 13^{\circ}\pm 11^{\circ},
\end{equation}
where the first error is statistical, the second is experimental systematics, and
the third is model uncertainty. 
The two standard deviation interval including the 
systematic and model uncertainties is $22^{\circ}<\phi_3<113^{\circ}$. 
The statistical significance of $CP$ violation for the combined measurement 
is 98\%. 

\section{Conclusion}

We report results of a measurement of the unitarity
triangle angle $\phi_3$ that uses a method based on a 
Dalitz plot analysis of the three-body $D$ decay in the process 
\bddsk.
The measurement of $\phi_3$ using this technique
was performed based on 253 fb$^{-1}$ data sample collected by 
the Belle detector. 
From the combination of \bdk\ and \bdsk\ modes, we obtain the 
value of 
$\phi_3=68^{\circ}\;^{+14^{\circ}}_{-15^{\circ}}\pm 13^{\circ}\pm 11^{\circ}$
(solution with $0<\phi_3<180^{\circ}$).
The first error is statistical, the second is experimental systematics and
the third is model uncertainty. 
The two standard deviation interval (including model and systematic 
uncertainties) is $22^{\circ}<\phi_3<113^{\circ}$.
The statistical significance of $CP$ violation for the combined 
measurement is 98\%. 
The method allows us to obtain a value of the 
amplitude ratio $r$, which can be used in other $\phi_3$ 
measurements. We obtain $r=0.21\pm 0.08\pm 0.03\pm 0.04$
for the \bdk\ mode and $r=0.12^{+0.16}_{-0.11}\pm 0.02\pm 0.04$ 
for the \bdsk\ mode.

\section*{Acknowledgments}

We are grateful to V.~Chernyak and M.~Gronau for fruitful 
discussions. 
   We thank the KEKB group for the excellent
   operation of the accelerator, the KEK Cryogenics
   group for the efficient operation of the solenoid,
   and the KEK computer group and the National Institute of Informatics
   for valuable computing and Super-SINET network support.
   We acknowledge support from the Ministry of Education,
   Culture, Sports, Science, and Technology of Japan
   and the Japan Society for the Promotion of Science;
   the Australian Research Council
   and the Australian Department of Education, Science and Training;
   the National Science Foundation of China under contract No.~10175071;
   the Department of Science and Technology of India;
   the BK21 program of the Ministry of Education of Korea
   and the CHEP SRC program of the Korea Science and Engineering 
Foundation;
   the Polish State Committee for Scientific Research
   under contract No.~2P03B 01324;
   the Ministry of Science and Technology of the Russian Federation;
   the Ministry of Education, Science and Sport of the Republic of 
Slovenia;
   the National Science Council and the Ministry of Education of Taiwan;
   and the U.S.\ Department of Energy.

\end{document}